%% file: main.tex
\documentclass[conference]{IEEEtran}
\IEEEoverridecommandlockouts
\usepackage{tikz}

\newcommand\submittedtext{%
  \footnotesize This work has been submitted to the IEEE for possible publication. Copyright may be transferred without notice, after which this version may no longer be accessible.}

\newcommand\submittednotice{%
\begin{tikzpicture}[remember picture,overlay]
\node[anchor=south,yshift=10pt] at (current page.south) {\fbox{\parbox{\dimexpr 0.65\textwidth-\fboxsep-\fboxrule\relax}{\submittedtext}}};
\end{tikzpicture}%
}
\usepackage{cite}
\usepackage{comment}
\usepackage{amsmath,amssymb,amsfonts}
\usepackage{textcomp}
\def\BibTeX{{\rm B\kern-.05em{\sc i\kern-.025em b}\kern-.08em
    T\kern-.1667em\lower.7ex\hbox{E}\kern-.125emX}}

\usepackage{quantikz}
\usepackage{float}
\usepackage{amsthm}
\theoremstyle{definition}

\usepackage{svg}
\usepackage{caption}
\usepackage{algorithm}
\usepackage{algorithmic}
\usepackage{graphicx}
\usepackage{xcolor}
\usepackage{bbm}
\usepackage{import}

\usepackage{acronym}
\acrodef{QSP}[QSP]{quantum state preparation}
\acrodef{MCX}[MCX]{multi-controlled NOT}
\acrodef{HHL}[HHL]{Harrow-Hassidim-Lloyd}
\acrodef{QFT}[QFT]{quantum Fourier transform}
\acrodef{QLS}[QLS]{quantum loading scheme}
\acrodef{NISQ}[NISQ]{noisy intermediate-scale quantum}
\acrodef{QRAM}[QRAM]{quantum random-access memory}
\acrodef{TSP}[TSP]{travelling salesman problem}

\makeatletter
\newcommand{\linebreakand}{%
  \end{@IEEEauthorhalign}
  \hfill\mbox{}\par
  \mbox{}\hfill\begin{@IEEEauthorhalign}
}
\makeatother

\usepackage{dcolumn}
\usepackage{bm}
\usepackage{mathtools}
\input{colors}

\newcommand{\vc}{\bm}
\newcommand{\mat}{\mathbf}

\renewcommand\bra[1]{{\langle{#1}|}}
\makeatletter
\renewcommand\ket[1]{%
  \@ifnextchar\bra{\k@t{#1}\!}{\k@t{#1}}%
}
\newcommand\k@t[1]{{|{#1}\rangle}}
\makeatother
\newcommand{\I}{\text{I}}
\newcommand{\Ftwo}[1]{\mathbbm{F}_2^{#1}}
\usepackage{soul}

\begin{document}

\title{
Sublinear Classical-to-Quantum Data Encoding using $n$-Toffoli Gates
\thanks{This study was funded by the QuantERA grant EQUIP via DFG project 491784278 and by the Federal Ministry for Economics and Climate Action (BMWK) via project ALQU and the Quantum Fellowship Program of DLR.}
}

\author{
\IEEEauthorblockN{Vittorio Pagni}
\IEEEauthorblockA{\textit{Institute of Software Technology} \\
\textit{German Aerospace Center (DLR)}\\
Sankt Augustin, Germany \\
\textit{University of Cologne}\\
Cologne, Germany\\
ORCID: 0009-0006-9753-3656}\\
\and
\IEEEauthorblockN{Gary Schmiedinghoff}
\IEEEauthorblockA{\textit{Institute of Software Technology} \\
\textit{German Aerospace Center (DLR)}\\
Sankt Augustin, Germany \\
ORCID: 0000-0003-2259-7365}
\and
\IEEEauthorblockN{Kevin Lively}
\IEEEauthorblockA{\textit{Institute of Software Technology} \\
\textit{German Aerospace Center (DLR)}\\
Sankt Augustin, Germany \\
ORCID: 0000-0003-2098-1494 }
\linebreakand
\IEEEauthorblockN{Michael Epping}
\IEEEauthorblockA{\textit{Institute of Software Technology} \\
\textit{German Aerospace Center (DLR)}\\
Sankt Augustin, Germany \\
ORCID: 0000-0003-0950-6801 }
\and
\IEEEauthorblockN{Michael Felderer}
\IEEEauthorblockA{\textit{Institute of Software Technology} \\
\textit{German Aerospace Center (DLR)}\\
Sankt Augustin, Germany \\
\textit{University of Cologne} \\
Cologne, Germany\\
ORCID: 0000-0003-3818-4442 }
}

\maketitle
\submittednotice
\begin{abstract}
Quantum state preparation, also known as encoding or embedding, is a crucial initial step in many quantum algorithms and often constrains theoretical quantum speedup in fields
such as quantum machine learning and linear equation solvers.
One common strategy is amplitude encoding, which 
embeds a classical input vector of size N=2\textsuperscript{n} in the amplitudes of an n-qubit register.
For arbitrary vectors, the circuit depth typically scales linearly with the input size N, rapidly becoming unfeasible on near-term hardware. We propose a general-purpose procedure with sublinear average depth in N, increasing the window of utility.

Our amplitude encoding method encodes arbitrary complex vectors of size N=2\textsuperscript{n} at any desired binary precision using a register with n qubits plus 2 ancillas and a sublinear number of multi-controlled NOT (MCX) gates, at the cost of a probabilistic success rate proportional to the sparsity of the encoded data.
The core idea of our procedure is to construct an isomorphism between target states and hypercube graphs, in which specific reflections correspond to MCX gates.
This reformulates the state preparation problem in terms of permutations and \emph{binary addition}. The use of MCX gates as fundamental operations makes this approach particularly suitable for quantum platforms such as \emph{ion traps} and \emph{neutral atom devices}. 
This geometrical perspective paves the way for more gate-efficient algorithms suitable for near-term hardware applications.
\end{abstract}


\section{Introduction}
Quantum computers can prepare and manipulate states within an exponentially large Hilbert space of dimension $N=2^n$ on an $n$-qubit register during computation. An exponential speedup is promised by algorithms such as the \ac{HHL} algorithm \cite{Harrow2009, b:Nielsen2012}, used for solving systems of linear equations, and the \ac{QFT} \cite{Coppersm2002, b:Nielsen2012}.
However, quantum-classical interfaces, i.e., loading classical data into the quantum register and reading out the quantum state, present a critical bottleneck for any algorithms aiming to operate on large input data.
Quantum machine learning, in particular, requires massive amounts of data to be embedded into the Hilbert space
\cite{b:Schuld2018}.
In the literature, the process of loading classical data into a quantum computer is variously called \emph{\ac{QSP}} \cite{Araujo2024}, \emph{encoding} \cite{Araujo2024}, \emph{quantum state loading} \cite{Hu2014a}, or \emph{quantum embedding} \cite{Phalak2024, Lloyd2020}.
Note that some of these terms have double meaning, such as encoding in the context of quantum data compression or embedding theory in quantum chemistry. 

Just as Holevo's theorem constrains the maximum information gain by readout of an $n$-qubit register \cite{Holevo1973}
likewise, an inefficient \ac{QSP} can eliminate any theoretical speedup of a quantum algorithm.
Particularly in the era of \ac{NISQ} devices requiring error mitigation, any reduction in circuit depth can mean the difference between useful and useless results \cite{Cai2023,Lively2024,Mangini2024}. 
Thus efficient embedding schemes are crucial on the quest for quantum utility. 

One proposed solution are specialized \ac{QRAM} \cite{Giovanne2008, Matteo2020} units to access stored quantum mechanical states and their superpositions. This approach presents major scaling challenges, some of which are architecture-specific \cite{Phalak2023, Jaques2023}, others that arise from more fundamental, physical limits \cite{Wang2024,Camacho2024}. Even without dedicated \ac{QRAM} hardware, one can optimize \ac{QSP} with optimal control \cite{Peng2017} or efficient gate-based ansatzes, which itself is treated as a circuit-based \ac{QRAM} in the literature. 
Different \ac{QSP} schemes have been proposed in the past, with various strengths and weaknesses. Often, there is a tradeoff between circuit depth and width, i.e., the number of non-simultaneously executable gates and qubits, respectively. 

For instance, basis encoding is used to embed classical data points $v_i$, $i\in[0,\dots,N-1]$ 
into quantum states of either the form $\sum_{i=0}^{N-1}|v_i\rangle / \sqrt{N}$ or  $\sum_{i=0}^{N-1}|i\rangle|v_i\rangle / \sqrt{N}$.
While the corresponding encoding circuits have sublinear depth $\mathcal{O}(\log(N))$, they typically require
lineary many qubits $\mathcal{O}(N)$ \cite{Hu2014a}.
On the other extreme, amplitude encoding of the form $\sum_{i=1}^{N}\sqrt{v_i}|i\rangle / \sum_i |v_i|$ only requires $\mathcal{O}(\log_2(N))$ qubits, in principle, but general-purpose amplitude-encoding circuits with $\mathcal{O}(\log(N))$ depth typically require $\mathcal{O}(N)$ ancillary qubits~\cite{Araujo2021}.
Angle-based embedding schemes, which use the classical data as gate rotation angles, require at least $\mathcal{O}(N)$ rotation gates \cite{stoudenmire2017}.

Many more proposals exist using at least $\mathcal{O}(N)$ CNOTs \cite{Plesch2011, Shende2006, Bergholm2005, Malvetti2021}.
Multiple works focus on optimized state preparation schemes, for instance by using low-rank approximations \cite{Araujo2024}, spectral decomposition methods \cite{Rosenkranz2024} or genetic algorithms \cite{Wright2024, Phalak2024}. Some proposal achieve high efficiency only for specific states, such as sparse states \cite{Mozafari2022, DeVeras2022, Malvetti2021, Gleinig2021}, uniform states \cite{Mozafari2021}, or those with matrix product state representation  \cite{GonzalezConde2024,Sato2025}. Some of these approaches allow for an adjustable tradeoff between circuit depth and width \cite{Araujo2023} or use probabilistic approaches \cite{Zhang2021, Park2019}.

Yet other works rely on state preparation with variational algorithms \cite{Ben-Dov2024, Marin-Sa2023, Nakaji2022, Zoufal2019}, which introduces significant preprocessing overhead for general states without efficient representations such as matrix product states. In particular, the open problem of barren plateaus may introduce $\mathcal{O}(N)$ optimization overhead.
A recent work encodes classical floating point values into quantum mechanical amplitudes to solve nonlinear partial differential equations \cite{Gaitan2024}, utilizing two ancilla qubits and $\mathcal{O}(\log(N))$ gates with a finite success probability. This is achieved through a binary expansion of the input vector and by performing a series of controlled rotations, which are tools that we also employ in this work. 

Here we show how the $L$-bit binary expansion of a generic, 
input vector of length $N$ can be loaded into the amplitudes of a non error-corrected quantum state of $n=\log_2(N)$ bits with 2 ancilla qubits by a sequence of \ac{MCX} gates, which are particularly suitable for NISQ platforms such as ion traps or neutral atom devices \cite{PhysRevA.101.032329}\cite{Yin_2020}\cite{goel2021nativemultiqubittoffoligates}. The logical flow of our approach is summarized in Figure \ref{eq:final_superposition}.

We 
begin by introducing the pre-processing Algorithm\,\ref{al:preprocessing}
in Sec.\,\ref{s:preprocessing} and describing the general procedure of the full Algorithm\,\ref{al:Mcx_ampl_enc}
in Sec.\,\ref{s:general_procedure}. The concrete properties of the \ac{MCX} circuit blocks are introduced in Sec.\,\ref{s:W_Implementation} and reformulated in the language of Kronecker decompositions to exploit binary operations for numerical efficiency and hypercube graphs \cite{mills1963cycles} for visualization and intuition.
We explain Subroutines\,\ref{al_sub:W_implementation} and \ref{al_sub:find_valid_nodes} that we use for finding efficient \ac{MCX} decompositions in Sec.\,\ref{s:tree_algorithm} and compare the performance for various input data against the Qiskit circuit in Sec.\,\ref{s:results}, where  for each encoding circuit we then analyze a shallower version (core) with a success probability that depends on the input and a longer version (full) which incorporates amplitude amplification to boost the success probability, before discussing the conclusions of our research in Sec.\,\ref{s:conclusion}. 

\begin{figure*}[tb]
    \centering
    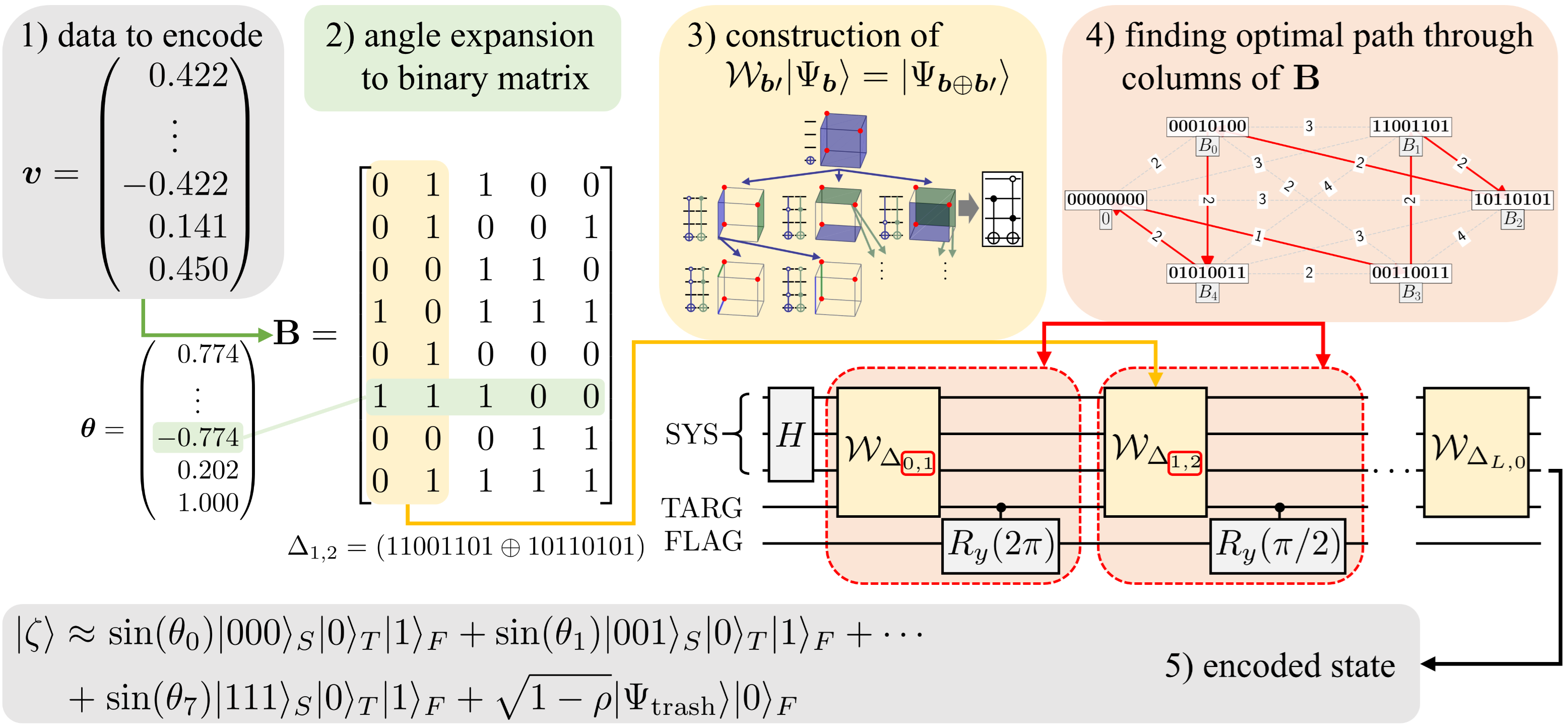
    \caption{
    Overview of the full algorithm. Input data (step 1) is transformed to renormalized angles whose binary expansion is stored in a matrix (step 2, see Sec.\,\ref{s:preprocessing}). Each column can be encoded with a circuit containing fully controlled \ac{MCX} (see Sec.\,\ref{s:general_procedure} and Sec.\,\ref{s:W_Implementation}), which we optimize using the tree algorithm (step 3, see Sec.\,\ref{s:tree_algorithm}). We further reduce the cost by constructing an optimized order of applying the encoding layers (step 4, see Fig.\,\ref{fig:solution_to_TSP}) to produce the encoded state (step 5, see \eqref{eq:final_superposition}).}
    \label{overview}
\end{figure*}

\section {Classical pre-processing} \label{s:preprocessing}
The first step consists of a classical pre-processing that, given a vector $\vc {v} \in \mathbb{R}^N$ and a bit precision $L \in \mathbb{N}$ as input, approximates the entries $\vc {v}$ with a $L$-bit binary expansion encoded in a binary matrix $\mat B\in \Ftwo{N\times L}$. This pre-processing scheme has already been presented in our previous work \cite{pagni2025}.
We use an intermediate rescaled angle vector $\vc{\theta}$, defined by 
\begin{equation}
   \theta_i = \arcsin \left( \frac{v_i}{\|\vc{v}\|_\infty} \right) \bigg/ \frac{\pi}{2}\in [-1,1]
\end{equation}
with $i \in \{0 \dots N-1\}$.
The signed L-bit binary representation (i.e. line 5 in Algorithm\,\ref{al:preprocessing}) of each $\theta_i$ is stored in the $i^{\text{th}}$ row $\vc{B}_{i,:}$ of $\mathbf{\mat B}$, where we use slice indexing notation ($:$) to indicate all elements in the row. Shortly, we will use the signed L-bit representation to encode an approximated vector $\vc w$, which converges to $\vc v$ as L increases. For complex input vectors, one encodes the modulus and phases separately \cite{pagni2025}.

\begin{algorithm}[H]
\caption{Classical Pre-Processing} \label{al:preprocessing}
\begin{algorithmic}[1] 
    \STATE \textbf{Input:} \begin {itemize} 
    \item real vector $\vc{v} \in \mathbb{R}^N$, $N=2^n$
    \item encoding precision $L \in \mathbb{N}$
    \end{itemize} 
    \STATE \textbf{Output:} binary matrix $\mat B\in \Ftwo{N\times L}$.

    \FOR{each $v_i \in \vc{v}$}
        \STATE $\theta_i \coloneqq \arcsin \left( v_i \big/ \|\vc{v}\|_\infty \right) \big/ \frac{\pi}{2}$
        \STATE set $B_{i,j} $ such that $\theta_i=(-1)^{B_{i,0}}\sum_{j=1}^{L-1} 2^{-j} B_{i,j} $
    \ENDFOR
    
    \RETURN $\mat{B}$
\end{algorithmic}
\end{algorithm}
\subsection{Example} \label{s:example}
If we choose $n=3$, $N=2^n=8$, $L=5$, and
\begin{equation}
    \vc{v}=\frac{1}{\sqrt{1265}}(15, 13, 10, -11, 12, -15, 5, 16) ~,
\end{equation}
 then the approximating vector is (here truncated to the second decimal digit) \begin{equation}
    \vc{w}=(0.43, 0.36, 0.26, -0.29, 0.33, -0.43, 0.13, 0.46) ~,
\end{equation}
the angle vector $\vc{\theta}$ and the binary matrix $\mat{B}$ are 
\begin{equation}
    \vc{\theta}=( 0.77, \ 0.60, \ 0.43, \ -0.48, \ 0.54, \ -0.77, \ 0.20, \ 1 )
\end{equation}
and
\begin{equation}\label{eq:example_B}
\mat{B} = \begin{bmatrix}
0 & 1 & 1 & 0 & 0 \\
0 & 1 & 0 & 0 & 1 \\
0 & 0 & 1 & 1 & 0 \\
1 & 0 & 1 & 1 & 1 \\
0 & 1 & 0 & 0 & 0 \\
1 & 1 & 1 & 0 & 0 \\
0 & 0 & 0 & 1 & 1 \\
0 & 1 & 1 & 1 & 1
\end{bmatrix} ~.
\end{equation}

\section{General Procedure} \label{s:general_procedure}

The full algorithm is sketched in Fig.\,\ref{overview}. In this Section, we will gradually build up the necessary concepts and conclude with the final Algorithm\,\ref{al:Mcx_ampl_enc}, which utilizes the subroutines that are discussed in Sec.\,\ref{s:tree_algorithm}.

\subsection{Encoding a single vector element}\label{IIIA}
The binary matrix $\mat B$ associated with the input vector $\vc{v}$ has $N$ rows, one for each entry in $\vc{v}$, and $L$ columns, which correspond to the $L$ bits in the binary representation of the $\vc{\theta}$ angle vector entries.
A key aspect of this encoding is that we can reconstruct each real entry $v_i$ by using $L$ controlled rotations with decreasing angle.
These rotation are either performed or skipped according to the $L$ bits in the corresponding row $\vc{B}_{i,:}$ of $\mathbf{\mat B}$. 

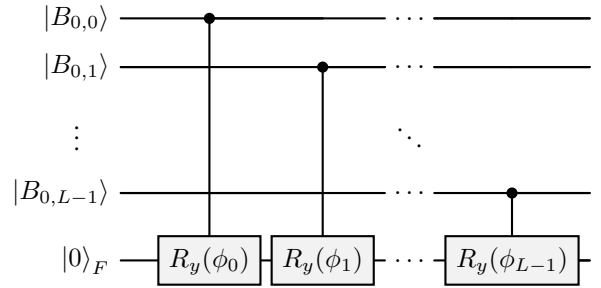
\begin{figure}[tb]
    \centering
    \input{controlled_rotations}
    \caption{An encoding layer for the first row of the matrix $\mat{B}$.}
    \label{mcx_encoder}
\end{figure}

Going back to the example, the first row is 
\begin{equation}
    \vc{B}_{0,:}=(0,1,1,0,0) \,.
\end{equation}
If we apply L sequential $R_y$ rotations conditioned on the entries of $\vc{B}_{0,:}$ with angles
\begin{align}
\label{phi_angles}
    \phi_l =
    \begin{cases}
        2 \pi &, l = 0 \\
        \pi/2^{l} &, l \in \{1, \dots, L-1\} \\
    \end{cases}
\end{align}
to a qubit whose initial state is $\ket{0}$, we obtain 
\begin{align*}
 \ket{\Psi_0}&=\prod_{l=0}^{L-1}\bigg(R_y(\phi_l)\bigg)^{B_{0,l}}\ket{0}\\
 &=\cos\left( \frac{1}{2}\sum_{l=0}^{L-1}B_{0,l}\cdot \phi_l \right) \ket{0} + \sin\left( \frac{1}{2}\sum_{l=0}^{L-1}B_{0,l}\cdot \phi_l \right) \ket{1}.
\end{align*}
Exploiting the definition of $\phi_i$ and  $\mathbf{B}$ we get 
\begin{align*}
\ket{\Psi_0} &= (-1)^{B_{0,0}} \cos\left( \arcsin\left(\left|\frac{v_0}{v_\infty}\right|\right) \right) \ket{0} \\
&+(-1)^{B_{0,0}} \sin\left( \arcsin\left(\left|\frac{v_0}{v_\infty}\right|\right) \right) \ket{1}
\end{align*}
and finally 
\begin{equation}
\begin{split}
    \ket{\Psi_0}=\text{sign}(v_0)\left(\sqrt{1-\bigg(\frac{v_0}{v_\infty}\bigg)^2}\ket{0}+\frac{v_0}{v_\infty}\ket{1}\right),
\end{split}
\end{equation}
thereby encoding the normalized value of $v_0$ into the amplitude of the $\ket{1}$ state. Since all $R_y$ commute, we can think of each $R_y(\phi_l)$ as encoding a specific precision level of the value $v_i/v_{\infty}$: the sign level $(-1)^B_{i,0}$, and each binary decimal level $2^{-j}B_{i,j}$. The corresponding circuit is shown in Fig.\,\ref{mcx_encoder}.
\subsection{Encoding an entire vector in parallel}
Now that we know how to exploit $R_y$ rotations in order to encode a single entry $v_i$ of the input vector, corresponding to a single row of $\mathbf{B}$, into the amplitude of the $\ket{1}$ state, we can parallelize this procedure over the whole vector using quantum superposition. The central idea is to consider the columns $\vc{B}_{:,l}, l \in \{0,\dots, L-1\}$ sequentially and 
at each step encode all entries of the column in parallel. 
The strategy we present here involves three registers:
\begin{itemize}
    \item a SYSTEM (S) register with $n$ qubits,
    \item a TARGET (T) register with one qubit,
    \item a FLAG (F) register with one qubit.
\end{itemize}
The algorithm begins with applying a layer of Hadamard gates to the SYSTEM register, creating the uniform superposition 
\begin{equation}
\label{initial_superposition}
    \ket{\Psi_{\vc 0}}_{S,T}:=\frac{1}{\sqrt N}\sum_{i=0}^{N-1}\ket{i}_S\ket{0}_T.
\end{equation}
Looking at this state, we can consider the basis state $\ket{i}$ on the SYSTEM register, ranging from $0$ to $N-1$, as an index for the state of the TARGET qubit. This allows us to see \eqref{initial_superposition} as encoding a specific binary vector of length $N$, which contains only zeros.
We now consider acting on this state with a unitary operator $\mathcal{W}_{\vc{b}} $ as a function of an arbitrary vector $\vc b \in \mathbb{F}_2^N$ such that 
it acts as
\begin{equation}
\label{W_definition_0}
\mathcal{W}_{\vc{b}}\ket{\Psi_{\vc 0}}=\ket{\Psi_{\vc {0} \oplus \vc{b}}}=\ket{\Psi_{\vc{b}}} \,,
\end{equation}
where $\ket{\Psi_{\vc b}}$ is the uniform  superposition
\begin{equation}
\label{arbitrary_superposition}
    \ket{\Psi_{\vc b}}_{S,T}:=\frac{1}{\sqrt N}\sum_{i=0}^{N-1}\ket{i}_S\ket{b_i}_T,
\end{equation}
in which the TARGET qubit is entangled with the SYSTEM register.
More generally given a state as in \eqref{arbitrary_superposition} and a binary vector $\vc b ' \in \mathbb{F}_2^N$ we want
\begin{equation}
\label{W_definition}
\mathcal{W}_{\vc{b'}}\ket{\Psi_{\vc b}}=\ket{\Psi_{\vc {b} \oplus \vc{b}'}}.
\end{equation}
For now we assume that such an operator $\mathcal{W}_{\vc{b'}}$ is given as a black box. 
We later describe how to implement the $\mathcal{W}$ operators in terms of \ac{MCX} gates in Sec.~\ref{s:W_Implementation}. Since, as it turns out, this decomposition is not unique, we explain our algorithm for finding efficient implementations in Sec.~\ref{s:tree_algorithm}.

When we act with a controlled $R_y(\phi)$ rotation that uses the qubit TARGET as control and the FLAG as target on the state in \eqref{arbitrary_superposition} tensored with the FLAG in the zero state, we get
\begin{align}
\label{controlled_RY_on_arbitrary_superposition}
    &CR_y(\phi_j,T\rightarrow F) \ket{\Psi_{\vc b}}_{S,T}\ket{0}_F\nonumber\\
    &= \frac{1}{\sqrt N}\sum_{i=0}^{N-1}\ket{i}_S\ket{b_i}_T (\cos(b_i \phi_j)\ket{0}_F+\sin(b_i \phi_j)\ket{1}_F) \nonumber \\
    &=\frac{1}{\sqrt N}\sum_{i=0}^{N-1}\ket{i}_S\ket{b_i}_T \bigg( R_y(b_i \phi_j)\ket{0}_F \bigg) .
\end{align}
From this equation we see how the binary values $b_i$ tell us which indices $i$ and therefore which amplitudes are affected by the given rotation $R_y(\phi_j)$ .

The final state we want to prepare is
\begin{align}
\label{eq:final_superposition}
\ket{\zeta}=&\frac{1}{\sqrt N}\sum_{i=0}^{N-1}\ket{i}_S\ket{ 0}_T \bigg( R_y\bigg(\sum_{j=0}^{L-1} \phi_{j} B_{i,j} \bigg)\ket{0}_F \bigg) \nonumber \\
=&\sqrt{\rho} \ket{\Psi_\text{good}}\ket{0}_T\ket{1}_F+\sqrt{1-\rho} \ket{\Psi_\text{bad}}\ket{0}_T\ket{0}_F
\end{align}
where 
\begin{equation}
\ket{\Psi_\text{good}}=\sum_{i=0}^{N-1} w_i \ket{i}_S.
\end{equation}
$\ket{\zeta}$ is a weighted superposition of the desired state that encodes the approximating vector $\vc{w}$, $\ket{\Psi_\text{good}}$, entangled with the state $\ket{1}$ on the FLAG register and of an undesired state $\ket{\Psi_\text{bad}}$, associated to the state $\ket{0}$ on the FLAG. By measuring the FLAG qubit, we know if the procedure succeeded, which happens with a probability given by the data density parameter \cite{pagni2025}
\begin{equation}
\label{density}
    \rho = \frac{1}{N} \sum_{i=0}^{N-1} \left(\frac{v_i}{\|\vc{v}\|_\infty}\right)^2 \in \left[\frac{1}{N}, 1\right] \,.
\end{equation}
If we measure zero on the FLAG qubit, we need to discard the state and repeat the encoding step. 
The success probability can be increased through amplitude amplification, but for near-term devices, where circuit depth is a bottleneck, it can be beneficial to instead use the shallow `core' encoding circuit with non-amplified success probability.

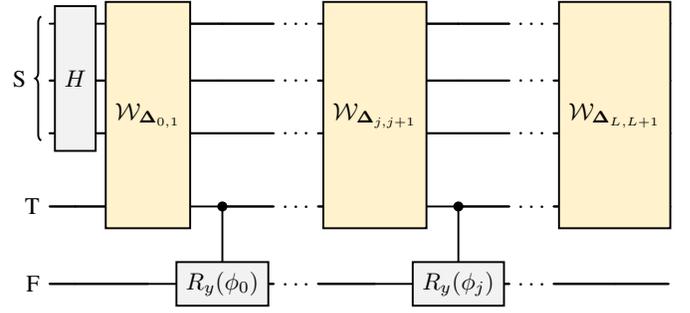
\begin{figure}
    \centering
    \scalebox{0.9}{\input{pictures/circuit_latex/algorithm_2_circuit}}
    \caption{Linear-path circuit that encodes the state $|\zeta\rangle$ starting from the initial state $|0\rangle$. It alternates the $L$ shift gates $\mathcal{W}_{\vc{\Delta}_{j,j+1}}$ and the controlled rotations $CR_y(\phi_j, T\to F)$ for $j\in \{0,1,\dots,L-1\}$ followed by the final disentangling shift $\mathcal{W}_{\vc{\Delta}_{L,L+1}}$. The costs $|\mathcal{W}_{\vc{\Delta}}|$ can be reduced by using a better permutation $\sigma$ of the $(\mathcal{W}, R_y)$ layers. The full protocol is described in Algorithm\,\ref{al:Mcx_ampl_enc}.
    }
    \label{fig:mcx_circuit}
\end{figure}

\begin{figure}
    \centering
    \includegraphics[width=1\columnwidth]{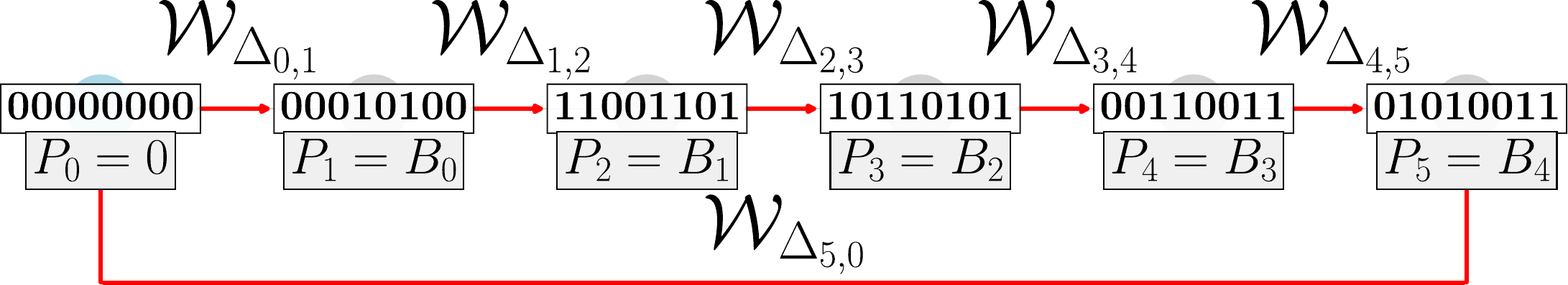}
    \caption{Example of the path P for the input binary matrix in  \eqref{eq:example_B}. The nodes are represented by the binary vectors, which are the columns of $\mat{B}$ plus the initial and final all-zero state, that need to be encoded into the superposition state of the three registers. }
    \label{fig:initial_tsp_path}
\end{figure}

In order to prepare the state $\ket{\zeta}$, we use the circuit in Fig.\,\ref{fig:mcx_circuit}.
We take the cost of preparing the superpositions of the  SYSTEM and TARGET registers with the circuit $\mathcal{W}_{\vc b}$ to be the number of \ac{MCX} gates in the decomposition found by Subroutine~\ref{al_sub:W_implementation}.
We denote the cost by $|\mathcal{W}_{\vc b}|$, motivated by the cardinality of the set of gates in the decomposition.
It turns out that the costs can be reduced by encoding the columns of $\mat B$ in a different order.
To that end, we introduce the path matrix $\mathbf{P} \in \Ftwo{N, L+2}$ containing a sequence of $L+2$ binary vectors 
\begin{equation}
\label{path_matrix}
  \mathbf P:=(\vc 0,\vc{B}_{:,0},\dots,\vc{B}_{:,L-1},\vc 0),
\end{equation}
 which is $\mathbf{B}$, padded with the all-zero column as the initial and final columns. We also use a padded $L+1$ angle vector 
\begin{equation}
    \vc\Phi:=(0,\phi_0,\dots,\phi_{L-1}) \,.
\end{equation}
The XOR of every two consecutive columns in $\mat P$ is
\begin{equation}
    \vc{\Delta}_{j,j+1}:=\vc P_{:,j} \oplus \vc P_{:,j+1}, j \in \{0,\dots,L\} \,.
\end{equation}
In the circuit, we alternate the $L$ shift operators $\mathcal{W}_{\vc{\Delta}_{j,j+1}}$ acting on the SYSTEM and TARGET registers with the $L$ controlled rotations $CR_y(\phi_j)$ targeting the FLAG qubit. 
Finally, we apply the shift operator $\mathcal{W}_{\vc{\Delta}_{L-1,L}}$ to disentangle the TARGET register from the others.
With the first and last components of $\mat P$ being zero vectors, we trivially have
\begin{gather}
    \mathcal{W}_{\Delta_{0,1}}=\mathcal{W}_{\vc 0 \oplus \vc {B}_{:,0}}=\mathcal{W}_{\vc {B}_{:,0}}\\
    \mathcal{W}_{\Delta_{L,L+1}}=\mathcal{W}_{\vc B_{:,L} \oplus \vc 0 }=\mathcal{W}_{\vc {B}_{:,L}} \,.
\end{gather}
This sequence of operators can be seen as a path along a graph whose nodes are the binary columns of $\mat P$ that can be encoded into the superposition in \eqref{arbitrary_superposition}, as shown in Fig.\,\ref{fig:initial_tsp_path}.
 
\begin{figure}
    \centering
    \includegraphics[width=1\columnwidth]{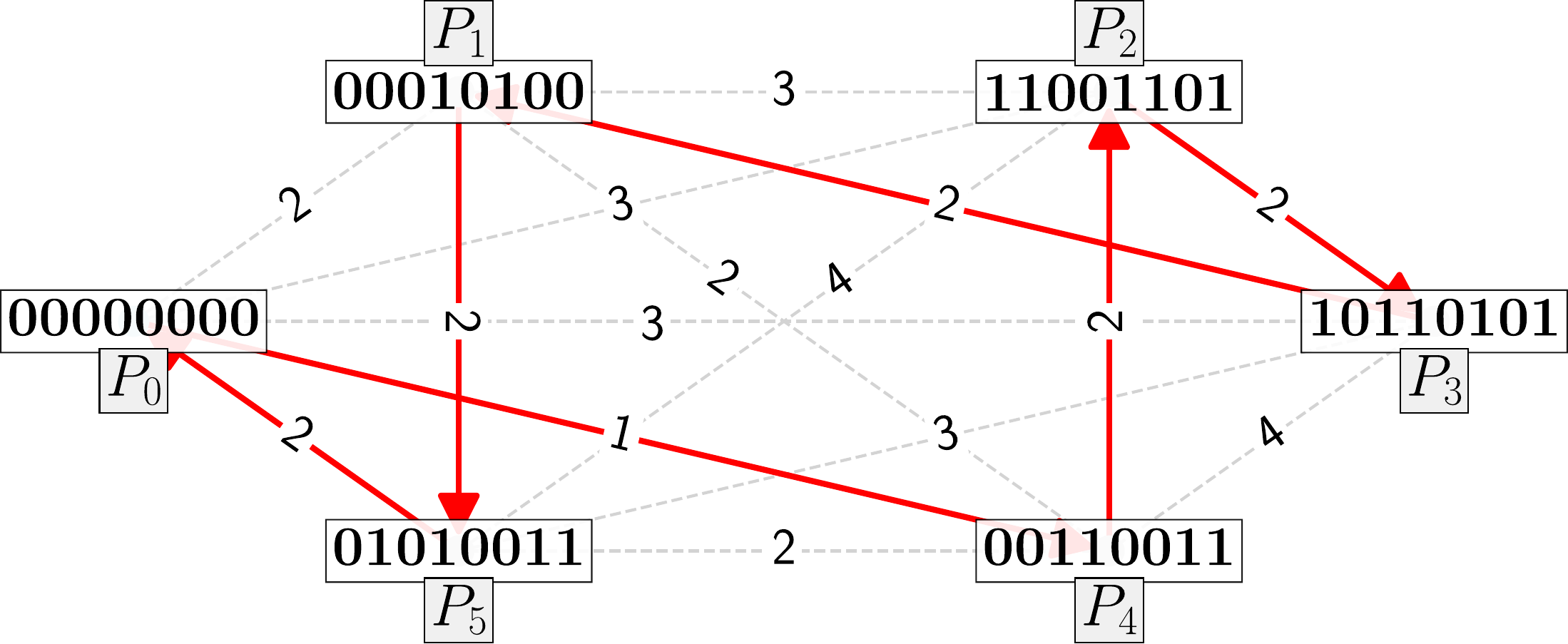}
    \caption{Complete graph representing all the possible encoding paths along the columns of $\mat P$, starting and ending with the all-zero columns $\vc P _{:,0}=\vc P _{:,L+1}$, Each edge  $(\vc P_{:,i},\vc P_{:,j})$ is associated with a vector $\vc \Delta_{i,j}=\vc P_{:,i} \oplus \vc P_{:,j}$ and with a cost $|\mathcal{W}_{\vc \Delta_{i,j}}| $, which is the amount of \ac{MCX} in the decomposition of the shift operator. We see in red the optimal path given by the permutation $\sigma = (4)(2)(3)(1)(5)$ for the example in \eqref{eq:example_B}. }
    \label{fig:solution_to_TSP}
\end{figure}

 At each application of $\mathcal{W}_{\vc{\Delta}{j,j+1}}$, the quantum state transitions from one node to the next in Fig.~\ref{fig:initial_tsp_path}, effectively shifting from a superposition state encoding column $\vc{P}_{:,j}$ to one encoding $\vc{P}_{:,j+1}$. Each controlled rotation $CR_y(\phi_j)$ adds the term $\vc P_{i,j+1}\phi_j$ for each state $\ket{i}$ of the SYSTEM in the superposition.
We can write the intermediate state of the SYSTEM as 
\begin{align}
\label{TARGET_superposition_jm}
  &\ket{\zeta_{j,m}}:=\frac{1}{\sqrt N}\sum_{i=0}^{N-1}\ket{i}_S\ket{ P_{i,j}}_T \bigg( R_y\bigg(\sum_{l=0}^{m} \phi_{l} P_{i,l} \bigg) \bigg)\ket{0}_F.
\end{align}
so that at the beginning of the $j^\text{th}$ encoding step we have the state $\ket{\zeta_{j,j}}$.
In this formalism, the initial state of the SYSTEM after the Hadamard layer described in \eqref{initial_superposition} can be seen as 
encoding the all-zero column $\vc P_{:,0}$ and having only $0\cdot P_{i,0}=0$ in the sum of angles, which corresponds to 
\begin{equation}
\ket{\zeta_{0,0}}=\frac{1}{\sqrt N}\sum_{i=0}^{N-1}\ket{i}_S\ket{ 0}_T\ket{0}_F=\ket{\Psi_{\vc 0}}_{S,T}\ket{0}_F.
 \end{equation}
Moreover, we can write our final target state from \eqref{eq:final_superposition} as
$
  \ket{\zeta}=  \ket{\zeta_{L+1,L}}.
$
During the j$^{\text{th}}$ encoding step we increment the first index by
$
    \mathcal{W}_{\vc{\Delta}_{j,j+1}} \ket{\zeta_{j,j}}=\ket{\zeta_{j+1,j}}
$
and then increment the second index by
$
     CR_y(\Phi_{j+1},T\rightarrow F) \ket{\zeta_{j+1,j}}=\ket{\zeta_{j+1,j+1}} \,.
$
As long as we begin and end with the all-zero columns $\vc{P}_{:,0}$ and $\vc{P}_{:,L+1}$, we can consider different permutations $\sigma$ of the intermediate $L$ columns of $\mat{P}$. These permutations correspond to the various paths in Fig.~\ref{fig:solution_to_TSP} and, due to the commutativity in the sum of angles, all yield the same final state $\ket{\zeta}$. Each edge of the path $(\vc P_{:,i},\vc P_{:,i})$ is assigned with a cost $|\mathcal{W}_{\vc \Delta_{i,j}}|$, which we take to be the amount of \ac{MCX} gates in $\mathcal{W}_{\vc \Delta_{i,j}}$.
Finding the optimal ordering $\sigma$ that minimizes the total cost 
\begin{equation}
\label{hypercube_distance}
    D(\sigma):=\sum_{j=0}^{L}|\mathcal{W}_{\vc{\Delta}_{\sigma(j),\sigma(j+1)}}|
\end{equation}
is equivalent to solving a \ac{TSP}.
The cost function $|\mathcal{W}_{\vc \Delta_{i,j}}|$ can be generalized
, for example by assigning weights to \ac{MCX} gates depending on the number of involved qubits.
 Algorithm~\ref{al:Mcx_ampl_enc} describes the whole procedure.
\begin{algorithm}[H]
\caption{\ac{MCX} Amplitude Encoder} \label{al:Mcx_ampl_enc}
\begin{algorithmic}[1]
\STATE \textbf{Input:}
\begin{itemize}
    \item binary encoding matrix $\mat{B} \in \mathbb{F}_2^{N \times L}$
    \item  $\ket{0}$-initialized quantum registers SYSTEM, TARGET, FLAG
\end{itemize}

\STATE \textbf{Output:}
\begin{equation*}
    \ket{\zeta}=\sqrt{\rho}
    \sum_{i=0}^{N-1} w_i \ket{i}_S \ket{0}_T\ket{1}_F+\sqrt{1-\rho} \ket{\Psi_\text{bad}}_S\ket{0}_T\ket{0}_F
\end{equation*}

  \STATE 
    $\ket{\zeta_{0,0}}=H^{\otimes n}\otimes I \otimes I\ket{0}=\frac{1}{\sqrt N}\sum_{i=0}^{N-1}\ket{i}_S\ket{0}_T\ket{0}_F$
\STATE $ \mathbf P=(\vc 0,\vc{B}_{:,0},\dots,\vc{B}_{:,L-1},\vc 0) $
\STATE call Subroutine\,\ref{al_sub:W_implementation} from Sec.\,\ref{s:tree_algorithm} to get costs $|\mathcal{W}_{\vc{\Delta}_{j,l}}|$
\STATE find best permutation $\sigma=\text{solveTSP}(\mat P, $\{$|\mathcal{W}_{\vc{\Delta}_{j,l}}|$\}$)$ \\
\hspace{1cm}where $\sigma(0)=0,\sigma(L+1)=L+1$
\STATE 
$\vc{\Delta}_{l,l+1}=\vc P_{:,\sigma(l)} \oplus \vc P_{:,\sigma(l+1)}, l \in \{0,\dots,L\} $
\FOR{$l \in \{0,\dots,L-1\}$}
\STATE 
 $\ket{\zeta_{l+1,l}}=\mathcal{W}_{\Delta_{l,l+1}}\ket{\zeta_{l,l}}$ 
\STATE 
 $\ket{\zeta_{l+1,l+1}} = CR_y(\Phi_{\sigma(l+1)}=\phi_l,T\rightarrow F)\ket{\zeta_{l+1,l}}$

 \ENDFOR
 \STATE disentangle TARGET register 
 $\mathcal{W}_{\Delta_{L,L+1}}\ket{\zeta_{L,L}}=\ket{\zeta_{L+1,L}}$ 
\RETURN 
$\ket{\zeta} = \ket{\zeta}_{L+1,L}$
\end{algorithmic}
\end{algorithm}

\section{Implementation of $\mathcal W$ with \ac{MCX} gates} \label{s:W_Implementation}

In this section, we explain how each $\mathcal{W}_{\vc b}$ can be implemented with \ac{MCX} gates.
The basic decomposition and important properties of the optimization problem are described in Sec.\,\ref{ss:W_properties}.
We then introduce two equivalent, but useful ways to view the problem: 
a formulation as finding the most efficient way to address vertices on a hypercube graph in Sec.\,\ref{ss:hypercube_view}, which provides a useful visual interpretation,
and a formulation as finding a minimal Kronecker decomposition in Sec.\,\ref{ss:kronecker_decomp_view}, which is particularly suited for numerical implementations with bit-wise operations. \\
We aim to construct a $\mathcal{W_{\vc b}}$ which has the properties in \eqref{W_definition_0} and \eqref{W_definition}. Looking at the equations  \eqref{initial_superposition} and \eqref{arbitrary_superposition}, we see that we can address every $b_i$, $i\in \{0,1,\dots,2^n-1\}$ in the superposition individually by means of one of the $2^n$ possible fully controlled MCX. As we will show in the following sections, using less than fully controlled MCX allows to simultaneously target specific subsets of the computational basis states, making the basis overcomplete and therefore allowing for shorter decompositions.
These pictures will aid us in formulating the optimization Subroutine\,\ref{al_sub:W_implementation} in Sec.\,\ref{s:tree_algorithm}, which finds a shallow \ac{MCX} circuit that implements $\mathcal{W}_{\vc b}$.

\subsection{Properties of $\mathcal{W}_{\vc b}$} \label{ss:W_properties}
\begin{figure}[tb] 
    \centering
    \includegraphics[width=1\columnwidth]{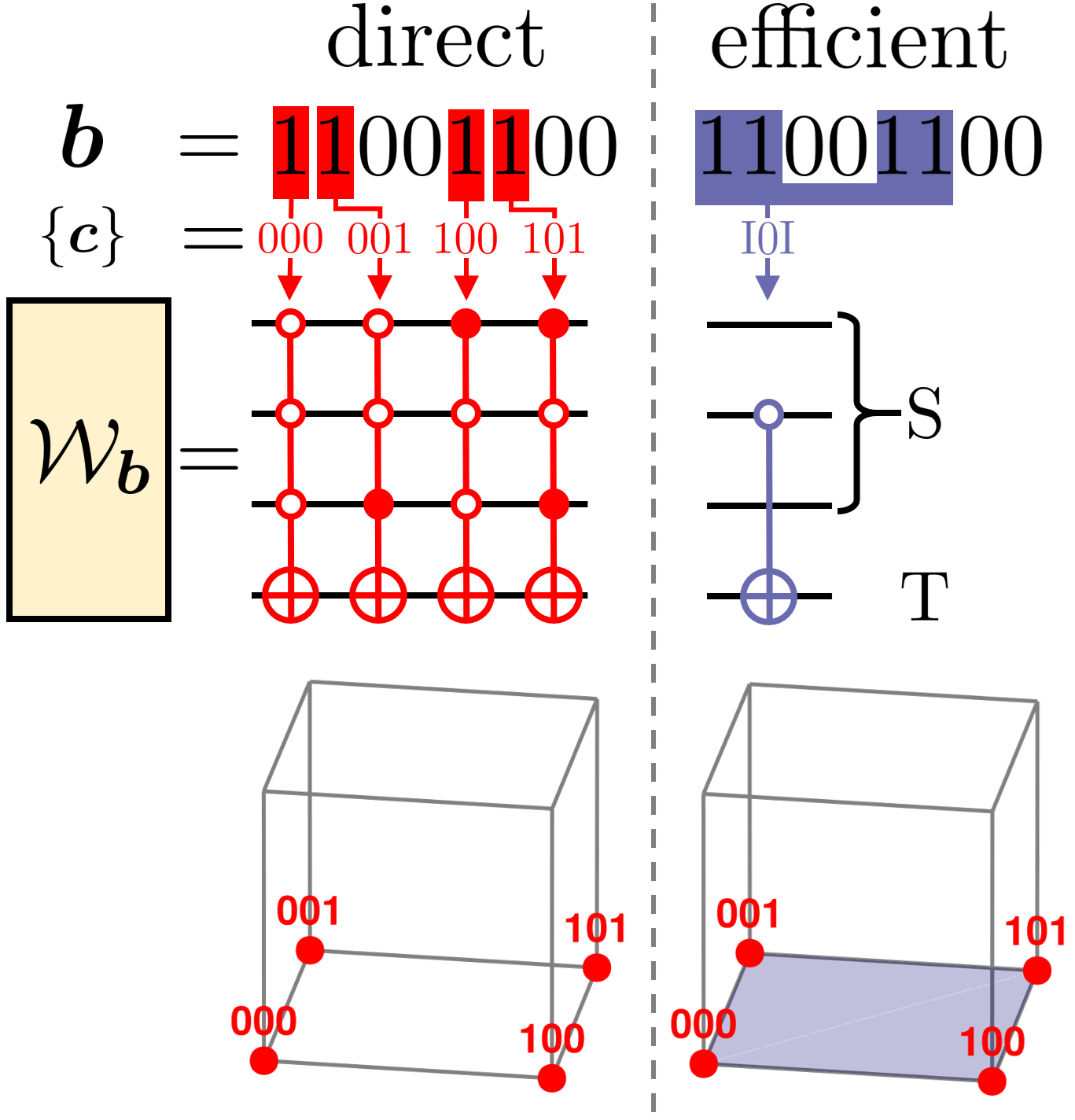}
    \caption{Two implementations of an exemplary $\mathcal{W}_{\vc b}$ circuit for the binary vector $\vc{b}=(1,1,0,0,1,1,0,0)$, which is shown as a bit string for brevity.
    The direct implementation (left, red) uses four fully controlled $\text{MCX}(\vc c_j,3)$ with control strings $\{\vc{c}_j\}=\{000, 001, 100, 101\}$ and
    the efficient implementation (right, blue) requires only one single-controlled $\text{MCX}(\I0\I,3)$ with $\vc c = \I0\I$. At the bottom we show the corresponding graph representations (see Sec.\,\ref{ss:hypercube_view}), where the control strings containing no (two) $\I$ symbols are visualized as nodes (faces) of the cube.}
    \label{fig:graph_language}
  \end{figure}

It turns out that each of the encoding circuits $\mathcal{W}_{\vc b}$ for a binary vector $\vc b$, which must fulfil \eqref{W_definition}, can be implemented by a product of \ac{MCX} gates
\begin{equation}
\mathcal{W}_{\vc b} = \prod_{j=0}^{M-1} \text{MCX}(\vc c_j) \label{eq:MCX_decomposition}
\end{equation}
which take the form
\begin{equation}\label{eq:MCX_projector}
    \text{MCX}(\vc c) = \mathbbm{P}_{\vc c,S} \otimes X_T + (\mathbbm{1}_S - \mathbbm{P}_{\vc c,S}) \otimes \mathbbm{1}_T \,,
\end{equation}
where $\vc c\in\{0,1,\I\}^n$ is the control string of the MCX, where we use the symbol $\I$ to indicate the uncontrolled qubit indices,
$\mathbbm{P}_{\vc c,S}$ is the projector to the corresponding subspace. Every $\I$ symbol in $\vc c$ indicates that $\mathbbm{P}_{\vc c}$ acts as the identity on the correspondent qubit. For example, for $n=3$  
\begin{equation}
    \mathbbm{P}_{00\vc\I,S}= \ket{00}\bra{00}\otimes \mathbbm{1}=\ket{000}\bra{000}+\ket{001}\bra{001}
\end{equation}
The most naive implementation of \eqref{eq:MCX_decomposition} uses fully controlled \ac{MCX}
and can be constructed by using the binary positions $\{ \vc{\nu} \,|\, b_{\vc{\nu}}=1 \}$ of each 1 in $\vc b$ as binary control strings $\{0,1\}^n$, see Fig.\,\ref{fig:graph_language}.
To reduce the number $M$ of \ac{MCX} gates, one must utilize those that only use a subset of the SYSTEM qubits as control,
which means trying to use $\vc c_j$ strings with the largest possible number of $I$ entries, denoted $\#_\I$.

The decomposition \eqref{eq:MCX_decomposition} is therefore not unique and finding the one that minimizes the amount $M$ of \ac{MCX} gates is in general an NP-hard problem, see Sec.\,\ref{ss:kronecker_decomp_view}. 
In fact, there are
\begin{equation}
    2^{(n-m)} \binom{n}{m}
\end{equation}
ways to choose $m$ control bits with either zero-control or one-control on $n$ bits.
The total amount of potential \ac{MCX} is equivalent to the number of possible strings $\vc c\in\{0,1,\I\}^n$,
\begin{equation}
\sum_{m=0}^{n} 2^{(n-m)} \binom{n}{m}=3^n.
\end{equation}
Finding an efficient decomposition of the operator \( \mathcal{W} \) as in \eqref{eq:MCX_decomposition}, with minimal \( M \), is equivalent to identifying a minimal set of control strings \( \{\vc{c}_k\} \) that fully specify the decomposition in terms of MCX gates. More generally, \( M \) may represent the total cost of implementing \( \mathcal{W} \), where each MCX gate in the decomposition is assigned a customizable cost, allowing for weighted optimization.
The order of strings $\vc{c}_k$ is irrelevant because all \ac{MCX} gates with the same target commute,
and all control strings in $\{\vc{c}_k\}$ must be unique for an optimal decomposition, because $\text{MCX}(\vc c)^2=\I$.
Any two \ac{MCX} whose control strings differ only in a single position can be joined
\begin{subequations} \label{eq:MCX_join}
\begin{align}
    \text{MCX}(\dots 0 \dots) \text{MCX}(\dots 1 \dots) &= \text{MCX}(\dots \I \dots) ~, \label{eq:MCX_join_01_to_I}  \\
    \text{MCX}(\dots 0 \dots) \text{MCX}(\dots \I \dots) &= \text{MCX}(\dots 1 \dots) ~,  \\
    \text{MCX}(\dots 1 \dots) \text{MCX}(\dots \I \dots) &= \text{MCX}(\dots 0 \dots) ~.
\end{align} 
\end{subequations}
In the same way, any \ac{MCX} can be split into two \acp{MCX}.
Using the relations \eqref{eq:MCX_join}, one can in principle manipulate the set of control strings $\{\vc{c}_k\}$ to find a more efficient decomposition \eqref{eq:MCX_decomposition}.
The properties of the \ac{MCX} gates directly transfer to the $\mathcal{W}$ operator 
\begin{subequations}
\begin{gather}
    \mathcal{W}^\dagger_{\vc b}=\mathcal{W}^{\phantom{\dagger}}_{\vc b}=\mathcal{W}^{-1}_{\vc b} \\
    \left[\mathcal{W}_{\vc b},\mathcal{W}_{\vc b'}\right]=0 ~,\\
    \mathcal{W}_{\vc b}\mathcal{W}_{\vc b'}=\mathcal{W}_{\vc b \oplus \vc b'} ~.
\end{gather}
\end{subequations}
\subsection{Hypercube graph representation} \label{ss:hypercube_view}
We now introduce an equivalent geometric description of the \ac{MCX} decomposition \eqref{eq:MCX_decomposition} in terms of a hypercube graph $G(\{\vc{c}_j |\, j \in \{0, \dots, M-1\}\})$, which is useful for visualization.
Generally, for any $n$-dimensional binary $\vc b$ 
we start by drawing the $n$-dimensional hypercube graph containing all nodes $\{0,1\}^n$.
In this graph, two nodes have an edge between them if their Hamming distance is 1. 
Next, we collect the binary positions $\{ \vc{\nu} \,|\, b_{\vc{\nu}}=1 \}$ of every 1 appearing in $\vc b$
and mark the corresponding nodes in the graph.
A simple example for $n=3$ can be seen on the left side of Fig.\,\ref{fig:graph_language}.

In this graph picture, each node corresponds to a control string $\vc c$ containing no identity symbols $\I$. 
A control string with a single $\I$ can be visualized as an edge and strings with two $\I$s correspond to faces (see right side of Fig.\,\ref{fig:graph_language}).
More generally, any control string $\vc c$ with $\I$s appearing at $\#_\I$ positions can be visualized as sub-hypercubes of dimension $n-\#_\I$.
The transformation \eqref{eq:MCX_join} on the \ac{MCX}-level correspond to joining subgraphs, e.g., two nodes into an edge, two edges into a face, or more generally, joining two $(\#_\I-1)$-dimensional sub-hypercubes into a $\#_\I$-dimensional sub-hypercube.
An example of this can be seen in Fig.\,\ref{fig:graph_language}, where four nodes are joined into a face, which corresponds to replacing four fully-controlled \ac{MCX} by a single \ac{MCX} with two $\I$ symbols.

\begin{figure}[tb] 
    \centering
    \includegraphics[width=0.68\columnwidth,trim={0mm 3mm 0mm 4mm},clip]{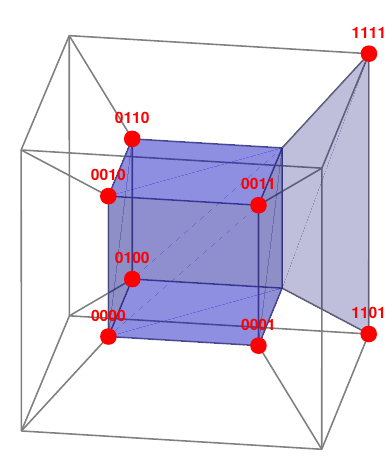}
    \caption{Visualization in $n=4$ dimensions of an efficient $\mathcal{W}_{\vc b}$ implementation for $\vc b = 1111101000000101$. The direct implementation uses eight fully-controlled \ac{MCX} with control strings $\{c_j\}_{\text{direct}}=\{ 0000, \allowbreak 0001, \allowbreak 0010, \allowbreak 0011, \allowbreak 0100, \allowbreak 0110, \allowbreak 1101, \allowbreak 1111 \}$. An efficient implementation uses only two \acp{MCX} $\{c_j\}_\text{efficient}=\{0\I\I\I, \allowbreak \I1\I1 \}$, i.e., the central cube $0\I\I\I$ and the right-rear face $\I1\I1$. This way, both gates include control on the SYSTEM states $0101$ and $0111$ which cancels out.
    }
    \label{fig:contraction_overlap}
\end{figure}

This visualization can be useful to get an intuitive understanding of efficient decompositions.
For instance, Fig.~\ref{fig:contraction_overlap} displays the graph of an efficient \ac{MCX} decomposition $\{c_j\}_\text{efficient}=\{0\I\I\I, \allowbreak \I1\I1 \}$
for $\vc b = 1111101000000101$ in $n=4$ dimensions
that exploits the involution $\text{MCX}(\vc c)^2=\I$
by including the same SYSTEM states in multiple control strings.
This solution may not be obvious when starting from 
$\{\vc{c}_j\}_\text{direct}=\{ \vc{\nu} \,|\, b_{\vc{\nu}}=1 \}$,
but is straightforward to see when drawing the graph for $\vc b$.
This example shows that sometimes, optimal solutions 
cannot be found by simply joining control strings to fewer control as in \eqref{eq:MCX_join_01_to_I},
i.e., only searching for sub-hypercubes embedded in the subgraph $G(\{c_j\}_\text{direct})$.

\begin{figure*}[tb] 
    \centering
    \includegraphics[width=\linewidth]{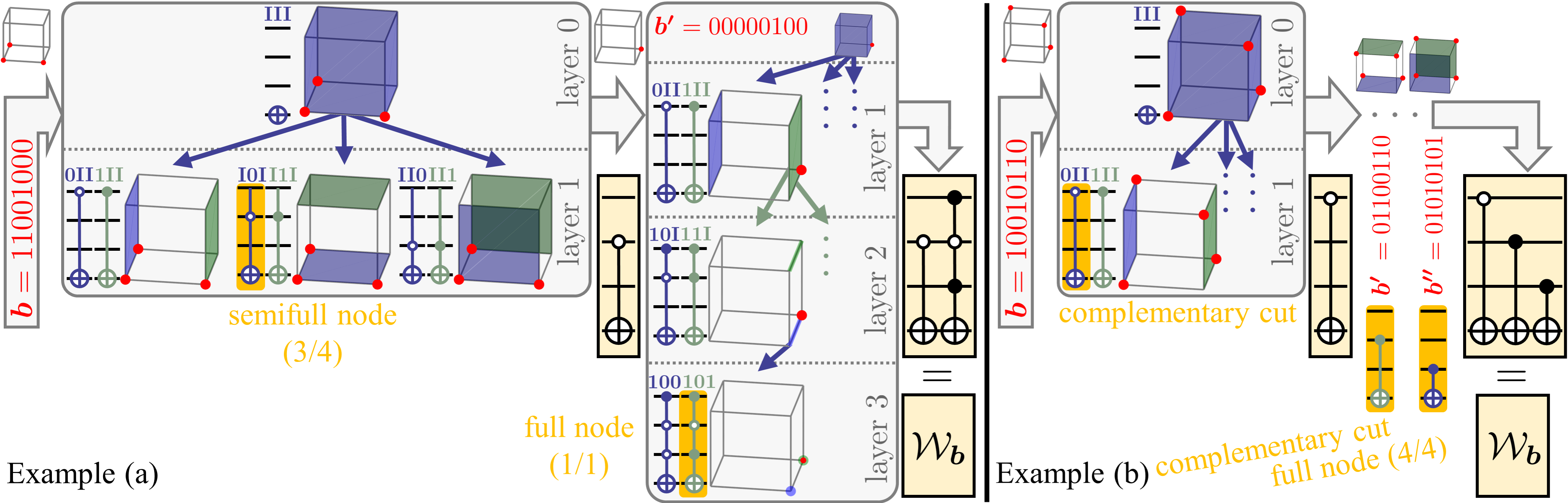}
    \caption{Visualization of Subroutine\,\ref{al_sub:W_implementation}, which produces a MCX-circuit $\mathcal{W}_{\vc b}$ for an input vector $\vc b$, on two exemplary inputs (a) $\vc b = 11001000$ and (b) $\vc b = 10010110$. In both cases, the tree is traversed from the top layer 0 (no control) downwards (layer $l$ contains \acp{MCX} with $l$ control qubits) until nodes of the tree are encountered that have either full or semi-full nodes or that have a complementary bi-partition. The tree algorithm is called repeatedly until all necessary \acp{MCX} for the circuit $\mathcal{W}_{\vc b}$ are found. Some branches and the last two trees of the second example are not displayed for brevity.}
    \label{fig:tree_algorithm}
\end{figure*}

\subsection{Kronecker decomposition representation} \label{ss:kronecker_decomp_view}
As a third option, we can reformulate the problem of finding an \ac{MCX}-decomposition \eqref{eq:MCX_decomposition} with minimal number $M$ of \ac{MCX}
into the problem of finding a minimal Kronecker decomposition, which allows for an efficient numerical implementation with bit-wise operations.
First, we associate each control string $\vc c \in \{0,1,\I\}^n$ with a binary vector $\vc b_{\vc c} \in \Ftwo{2^n}$ via the following mapping 
\begin{equation}
   \vc b_{\vc c}:=\bigotimes_{i=0}^{n-1} \vc t(c_i)
, \quad
\vc{t}(c_i) =
\left\{
    \begin{aligned}
        (1,0) &\qquad \text{if } c_i=0\\
        (0,1) &\qquad \text{if } c_i=1\\
        (1,1) &\qquad \text{if } c_i=\I
    \end{aligned}
\right.
\end{equation}
expresses the binary vector $\vc b_{\vc c}$ as a tensor product of 2-dimensional binary vectors in $\{(1,0),(0,1),(1,1)\}$.
Every tensor product corresponds to a single \ac{MCX} as
\begin{equation}
    \mathcal{W}_{\vc{b}_{\vc c}}=\text{MCX}(\vc c) ~.
\end{equation}
A general binary vector $\vc b \in \mathbb{F}_2^N$ can be decomposed into Kronecker products
\begin{equation}
\label{kronecker_decomposition}
    \vc b= \bigoplus_{j=0}^{M-1} \vc b_{\vc c_j} ~,
\end{equation}
which is equivalent to the \ac{MCX} decomposition \eqref{eq:MCX_decomposition}.
Thus, finding a minimal \ac{MCX} decomposition of $\mathcal{W}_{\vc b}$ is equivalent to finding the shortest Kronecker decomposition,
which is known to be NP-hard \cite{boolean_matrix_factor,fast_efficient_BMF}.
The advantage of this formulation is that the joining and splitting operations \eqref{eq:MCX_join} can be expressed as binary additions (XOR operations) on the tensor products $\vc b_{\vc c}$.
Thus, in Sec.\,\ref{s:tree_algorithm} we will formulate our algorithm in terms of Kronecker decompositions.

\section{Algorithm for efficient construction of $\mathcal{W}_{\vc b}$} \label{s:tree_algorithm}

The full Algorithm\,\ref{al:Mcx_ampl_enc} uses the following Subroutine.
\captionsetup[algorithm]{name=Subroutine}
\setcounter{algorithm}{0}
\begin{algorithm}[H]
\caption{construct\_$\mathcal{W}_{\vc{b}}$ (Kronecker Decomposition)}
\begin{algorithmic}
\label{al_sub:W_implementation}
\STATE \textbf{Input:} binary vector $\vc  b \in \Ftwo{N},N=2^n$
\STATE \textbf{Output:} control strings $C_{\mathcal{W}_{\vc{b}}} \subset \{0,1,\I\}^n$ for an efficient\\
\STATE \hspace{12mm} decomposition $\vc b= \bigoplus_{j=0}^{M-1} \vc b_{\vc c_j}$, see \eqref{kronecker_decomposition} 
\STATE $C_{\mathcal{W}_{\vc{b}}}$ = $\emptyset$
\WHILE{$\vc{b}$ != $\vc{0}$}
    \STATE $C_\text{candidates}$ = findNextControlStrings($\vc{b}$)
    \STATE $C_\text{selected}$ = findMaxIndependentSet($C_\text{candidates}$)
    \STATE $\vc{b}$ = $\vc{b} \bigoplus \{\vc{b}_{\vc{c}} \,|\, \vc{c} \in C_\text{selected} \}$ 
    \STATE $C_{\mathcal{W}_{\vc{b}}}$ = $C_{\mathcal{W}_{\vc{b}}} \, \cup \, C_\text{selected}$
\ENDWHILE
\RETURN $C_{\mathcal{W}_{\vc{b}}}$
\end{algorithmic}
\end{algorithm}
For a given $\vc{b}$, this subroutine returns control strings $C_{\mathcal{W}_{\vc{b}}} \subset \{0,1,\I\}^n$ for an efficient $\mathcal{W}_{\vc b}$ circuit, as illustrated in Fig.\,\ref{fig:tree_algorithm}.
For a visual understanding of the subroutine, we can sort all control strings $\{0,1,\I\}^n$ into a tree,
where the layer index $l$ corresponds to how many control qubits (how many symbols 0 and 1) are used.
The root of the tree contains the string $\I^{\otimes n}$ and sits in layer 0.
For each control string in a node, we create a new branch for every $\I$ symbol, which leads to a node containing the bi-partition $(\vc{c}_0, \vc{c}_1)$, where the selected $\I$ is replaced by either 0 or 1. 
The best candidates for controls strings
are found by calling Subroutine\,\ref{al_sub:find_valid_nodes}, from which
Subroutine\,\ref{al_sub:W_implementation} selects
a maximum independent set $C_{\text{selected}} = \{\vc{c} \,|\, \vc{b}_{\vc{c}} \mathbin{\&} \vc{b}_{\vc{c}'} = \vc{0}\}$ with a greedy algorithm.
Each selected control string $\vc{c}$ is added to the solution $\mathcal{W}_{\vc{b}}$ and the binary vector is transformed to $\vc{b} \to \vc{b} \oplus \vc{b}_{\vc{c}}$. This is repeated until the binary vector is transformed to $\vc 0$. 

Subroutine\,\ref{al_sub:find_valid_nodes} finds the lowest tree layer with valid nodes and returns that layer's most promising control strings $C_\text{candidates}$.
\begin{algorithm}[H]
\caption{findNextControlStrings}
\begin{algorithmic} 
\label{al_sub:find_valid_nodes}
\STATE \textbf{Input:} binary vector $\vc  b \in \Ftwo{N},N=2^n$
\STATE \textbf{Output:} set $C_\text{candidates} \subset \{0,1,\I\}^n$ of control strings
\STATE \hspace{12mm} that can be applied next in Subroutine\,\ref{al_sub:W_implementation}
\STATE  currentLayer = \{root\} = $\{\I^{\otimes n}\}$
\WHILE{true}
    \STATE currentLayer = doCutsToGetNextLayer(currentLayer)
    \IF{hasFullNodes(currentLayer)}
       \RETURN $ \{ \vc{c} \in \text{currentLayer} \,|\, F_{\vc{b}, \vc{c}} = 1\}$
    \ELSIF{hasSemiFullNodes(currentLayer)}
        \RETURN $ \{ \vc{c} \in \text{currentLayer} \,|\, F_{\vc{b}, \vc{c}} \geq 3/4\}$
    \ELSIF{hasComplNodes(currentLayer)}
        \RETURN $\{ \vc{c}_0 \,|\, (\vc{c}_0, \vc{c}_1) \in \text{complNodes(currentLayer)} \}$
    \ENDIF
\ENDWHILE
\end{algorithmic}
\end{algorithm}
Valid nodes can appear in three types: full nodes, semi-full nodes and complementary cuts.
Full nodes are prioritized over semi-full nodes, which are preferred over complementary cuts. 
They are defined by their fullness
\begin{equation}
    F_{\vc{b}, \vc{c}} = \frac{\#_1 ( \vc{b}_{\vc c} \mathbin{\&} \vc{b} )} { \#_1 (\vc{b}_{\vc c})} \in [0,1],
\end{equation} with the bitwise AND operator $\mathbin{\&}$.
It is desirable to select a $\vc{c}$ with many $\I$'s (low layer in tree) and high $F_{\vc{b}, \vc{c}}$, because this removes many 1's from $\vc{b}$ with a single \ac{MCX}.
We define a \emph{full} node to have $F_{\vc{b}, \vc{c}}=1$ and a \emph{semi-full} node to have $F_{\vc{b}, \vc{c}}\geq 3/4$, see example (a) in Fig.\,\ref{fig:tree_algorithm}. A \emph{complementary cut} is a bi-partition $(\vc{c}_0, \vc{c}_1)$ 
in whose selected subspaces the 1-positions $\{ \vc{\nu} \,|\, b_{\vc{\nu}}=1 \}$ of $\vc b$ are complementary.
This is most-easily understood graphically, see example (b) in Fig.\,\ref{fig:tree_algorithm}.
Formally, a cut on layer $l$ is complementary if
$\vc{b}_{\vc{c_0}} \mathbin{\&} \bar{\vc{b}}$ is identical to $(\vc{b}_{\vc{c_1}} \mathbin{\&} \vc{b})  \ll l$ with the bit-shift operator $\ll$ and the complementary binary $\bar{\vc b}$ of $\vc b$.
When no (semi-)full node is found in a layer, but it has a complementary cut $(\vc{c}_0, \vc{c}_1)$, it is still beneficial to add either
$\vc{c}_0$ or $\vc{c}_1$ to the solution, because this creates additional full nodes for the next steps.

\begin{figure*}[tb]
    \centering
    \includegraphics[width=1\linewidth,trim={5mm 7mm 5mm 3mm},clip]{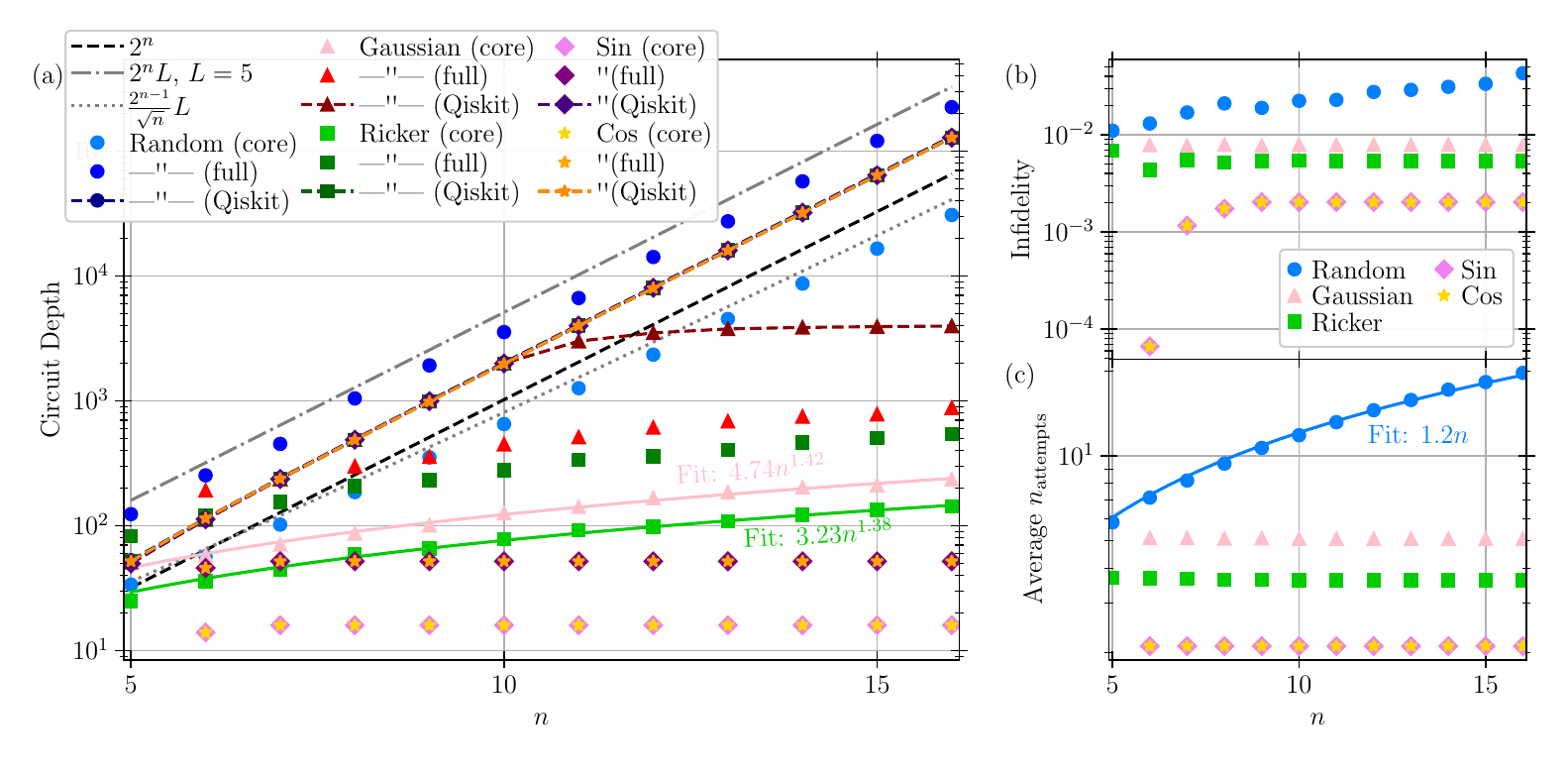}
    \caption{Scaling of the MCX encoder algorithm circuit depth (left) with (full) and without (core) subsequent amplitude amplification, infidelity (top right) and average number of necessary attempts for a successful encoding without amplitude amplification (bottom right), for different input classes with $n \in [5, 16]$ and $L = 5$.
    The data spans common test functions, including:
    (i) a random vector with entries drawn i.i.d.\ from a standard normal distribution (before normalization), 
    (ii) a Ricker wavelet and a Gaussian function, both with $\sigma = 1$, $\mu = 0$, sampled in the interval $[-3, 3]$, and 
    (iii) the functions $\sin(x)$ and $\cos(x)$ sampled in the interval $[0, 2\pi]$. The $\text{---''---}$ character in the legend refers to the previously mentioned input type.
   The depths are compared to equivalent Qiskit encoding circuits transpiled to the gate basis {\texttt{Rx}, \texttt{Ry}, \texttt{Rz}, \texttt{CNOT}} using the highest optimization level (level 3). All results from Qiskit overlap, showing exponential scaling greater than $2^n$, except for the circuit encoding the Gaussian, whose depth saturates at around 1300 at $n=11$. Solid lines indicate fits to the numerical data, while the dashed lines are included to aid visualization.
}
    \label{fig:mcx_scaling_comparison}
\end{figure*}


\begin{figure}[H]
    \centering
    \includegraphics[width=1\columnwidth,trim={5mm 5mm 5mm 5mm},clip]{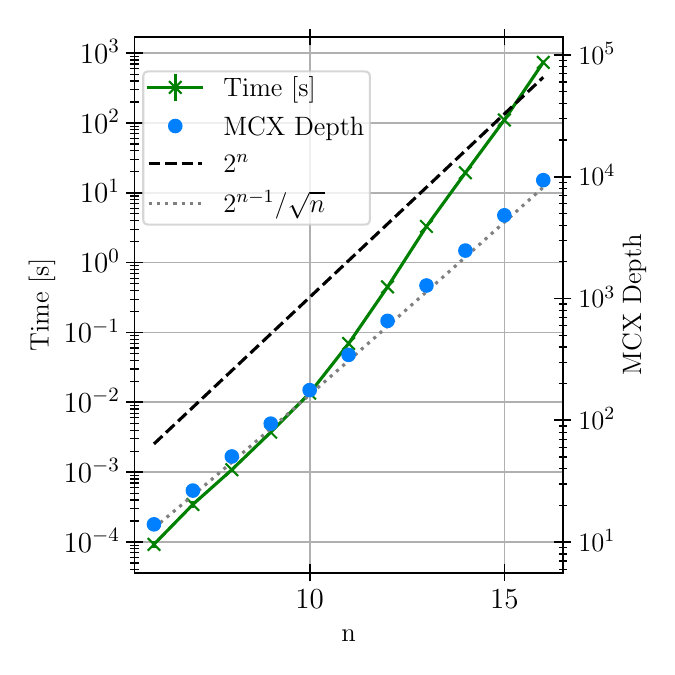}
    \caption{Single core performance of the encoding subroutine for a single binary vector. The left y axis shows the walltime to decompose an arbitrary vector of size $2^n$ on a single Intel Xeon Platinum 8280 CPU core for $n$ number of qubits on the x axis. Results are averaged over 10 random binary vectors at each $n$. 
    Standard deviation of the time results are not visible at this scale. The right y axis shows the resultant depth of the MCX gate decomposition. The $2^n$ and $2^n/\sqrt{n}$ lines are plotted on the same scale as the MCX depth. This code can be straightforwardly parallelized across the tree search.
    }
    \label{fig:single_bin_vec_scaling}
\end{figure}

\section{results} \label{s:results}
To evaluate the performance and scalability of our MCX encoding algorithm, we conduct a 
benchmarking analysis across a range of input vector classes and 
sizes. In particular, we analyze how the circuit depth, infidelity, and success probability of the core circuits scale with the number of qubits.
The considered inputs span both structured and unstructured data, allowing us to highlight the advantages of our approach in practical settings. Our test set includes:
\begin{itemize}
    \item  a vector randomly sampled with i.i.d. standard normal entries and then normalized to one, which is equivalent to uniformly sampling from the hypersphere $\mathcal{S}_{N-1}$ \cite{pagni2025}.
\item  a Ricker wavelet and a Gaussian function, both sampled in the interval $[-3, 3]$ with $\sigma = 1$, $\mu = 0$, and 
\item  the functions $\sin(x)$ and $\cos(x)$ sampled uniformly in the interval $[0, 2\pi]$. 
\end{itemize}
For each input type we compare the circuit depth 
to 
the correspondent Qiskit 
circuit, after transpiling 
the latter into the basis set \{\texttt{Rx}, \texttt{Ry}, \texttt{Rz}, \texttt{CNOT}\} using the highest available optimization level (level 3), with an all-to-all connected geometry \cite{qiskit_state_prep,qiskit2024}.

Figure~\ref{fig:mcx_scaling_comparison} 
shows the results of this analysis.
We choose the precision $L$ to be 5. This comparison sheds light on the distinct scaling regimes estimated numerically and highlights the efficiency gains achievable with the \ac{MCX} encoder, particularly for data exhibiting regularity, where our approach provides an exponential improvement over Qiskit for both the full and the core versions, showing a poly-log scaling in $N$ for both the Gaussian and Ricker wavelet data while both saturating to the same relatively small constants for $\cos(x)$ and $\sin(x)$, namely 16 (core) and 52 (full), with a success probability of roughly $50 \%$ for the former. While the full \ac{MCX} circuit has 
depth $\approx 2^n L$ for unstructured data---worse than Qiskit in this case---the \emph{core} \ac{MCX} circuit demonstrates a better scaling of $\approx \frac{2^{n-1} L}{\sqrt{n}}$, which significantly outperforms Qiskit for any fixed $L$ and extends the window of utility. The infidelity, which can be arbitrarily decreased by choosing greater values of $L$, remains constant and below 1\% for all examples of structured data, while it slowly increases for unstructured data, reaching up to 4\% for $n = 16$. The average number of attempts, $n_\text{attempts}$, needed to successfully prepare the state using the core MCX circuit remains constant and below 5 for all the structured inputs we considered, and scales linearly with the number of qubits for unstructured data, in accordance with the asymptotic behavior of $\frac{1}{\rho}$ (see (16)) described in~\cite{pagni2025}.

Figure \ref{fig:single_bin_vec_scaling} shows the efficiency of our implementation of the central computational bottleneck of Algorithm\,\ref{al:Mcx_ampl_enc}. This consists of Subroutines\,\ref{al_sub:W_implementation} and \ref{al_sub:find_valid_nodes}, which find the Kronecker decomposition of arbitrary binary vectors of size $2^n$. By working directly with 64-bit unsigned integer data types we were able to optimize the implementation to where the bottleneck operations are bitwise operators. Figure \ref{fig:single_bin_vec_scaling} shows how the walltime for the decomposition of single random bitvectors on a single CPU core scales as roughly $2^n$, until reaching bitvectors of size $1024$ or $16$ UInt64 entries, where the scaling rules change, likely due to the efficiency of CPU operations over the UInt64 vectors. This code can be straightforwardly parallelized to further improve the performance.

\section{Conclusion \& Outlook} \label{s:conclusion}
In this work, we introduced an algorithm to embed an $L$-bit binary approximation of $N$ classical real values
into a quantum register of $n=\log_2(N)$ qubits via amplitude encoding, using $L$ rotations and a series of \ac{MCX} operations and 2 ancilla qubits.
Our algorithm first builds a binary matrix $\mat B \in \mathbb{F}_2^{N,L}$ which contains the binary expansions of the input values,
with each one of the columns corresponding to a power of two in the expansion.
It then solves a \ac{TSP} to find the optimal path in which to encode the $L$ columns.
The cost of the path depends on the required number of \acp{MCX} to build the difference vector between consecutively visited columns.
To find an efficient \ac{MCX} decomposition, we use a tree algorithm, which reformulates the problem as finding the minimal Kronecker decomposition.
This allows us to use binary operations in the numerical implementation.
In addition, we introduced an isomorphism to a hypercube graph,
which enables useful visualizations of our algorithm and helps in building an intuitive understanding of good solutions.

We analyzed the performance of our algorithm on several examples of structured input data from the literature, and compared the circuit depth with that of equivalent Qiskit encoding circuits transpiled to the gate set \{\texttt{Rx}, \texttt{Ry}, \texttt{Rz}, \texttt{CNOT}\} at the highest available optimization level (level 3). 
For each encoding circuit, we analyzed a shallower version (core), with a success probability that depends on the input density $\rho$ in  \eqref{density}, and a longer version (full), which incorporates amplitude amplification to boost the success probability.
For both versions, we present numerical evidence for a scaling of the circuit depth that is $\mathcal{O}(n^\alpha L)$, $\alpha \approx 1.4$ for common structured inputs such as Gaussian or Ricker wavelet distributions, and $\mathcal{O}(L)$ for functions like $\cos(x)$ and $\sin(x)$, providing an exponential improvement over Qiskit. The success probability is constant in $N$ and exceeds $20\%$ in all cases.
For unstructured data, while the full \ac{MCX} circuit has 
depth $\approx 2^n L$, which is worse than Qiskit, the \emph{core} \ac{MCX} circuit shows improved scaling of $\approx \frac{2^{n-1} L}{\sqrt{n}}$, significantly outperforming Qiskit for any fixed $L$, with success probability scaling as $\frac{1}{n}$.

Even if not fully deterministic, we consider the core version of each encoding circuit to be of interest for NISQ devices, where a shorter circuit reduces noise and errors at the cost of rejecting some trials. This is particularly relevant given that the average number of required attempts is independent of $N$ for structured inputs and scales only slowly, as $\log N$, for unstructured input.\\
In the future, the algorithm can be further improved.
The current low-level implementation is highly optimized using binary operations, but algorithmic improvements are possible.
For instance, one can adapt Subroutines\,\ref{al_sub:W_implementation} and \ref{al_sub:find_valid_nodes} to be more discriminating between semi-full nodes by using the precise fullness values to determine an optimal selection of nodes from the maximum independent sets. Additionally, on a more fundamental level, one can choose a different basis than the binary one given by $\mat{B}$, such as the Hadamard–Walsh basis, which we explore in ongoing work.
Finally, 
one can include \acp{MCX} which target the SYSTEM register, 
which in the hypercube picture corresponds to reflections.
Such operations sacrifice the commutativity of the \acp{MCX}, but vastly increase the search space of solutions. 

\bibliographystyle{IEEEtran}
\bibliography{main}

\end{document}

%% file: colors.tex
\definecolor{gate_color_H}{rgb}{0.949,0.949,0.949} 
\definecolor{gate_color_W}{rgb}{1,0.949,0.80}
\definecolor{gate_color_crot}{rgb}{0.949,0.949,0.949} 

%% file: pictures/overview.pdf_tex
\begingroup%
  \makeatletter%
  \providecommand\color[2][]{%
    \errmessage{(Inkscape) Color is used for the text in Inkscape, but the package 'color.sty' is not loaded}%
    \renewcommand\color[2][]{}%
  }%
  \providecommand\transparent[1]{%
    \errmessage{(Inkscape) Transparency is used (non-zero) for the text in Inkscape, but the package 'transparent.sty' is not loaded}%
    \renewcommand\transparent[1]{}%
  }%
  \providecommand\rotatebox[2]{#2}%
  \newcommand*\fsize{\dimexpr\f@size pt\relax}%
  \newcommand*\lineheight[1]{\fontsize{\fsize}{#1\fsize}\selectfont}%
  \ifx\svgwidth\undefined%
    \setlength{\unitlength}{513.99920005bp}%
    \ifx\svgscale\undefined%
      \relax%
    \else%
      \setlength{\unitlength}{\unitlength * \real{\svgscale}}%
    \fi%
  \else%
    \setlength{\unitlength}{\svgwidth}%
  \fi%
  \global\let\svgwidth\undefined%
  \global\let\svgscale\undefined%
  \makeatother%
  \begin{picture}(1,0.44007)%
    \lineheight{1}%
    \setlength\tabcolsep{0pt}%
    \put(0,0){\includegraphics[width=\unitlength,page=1]{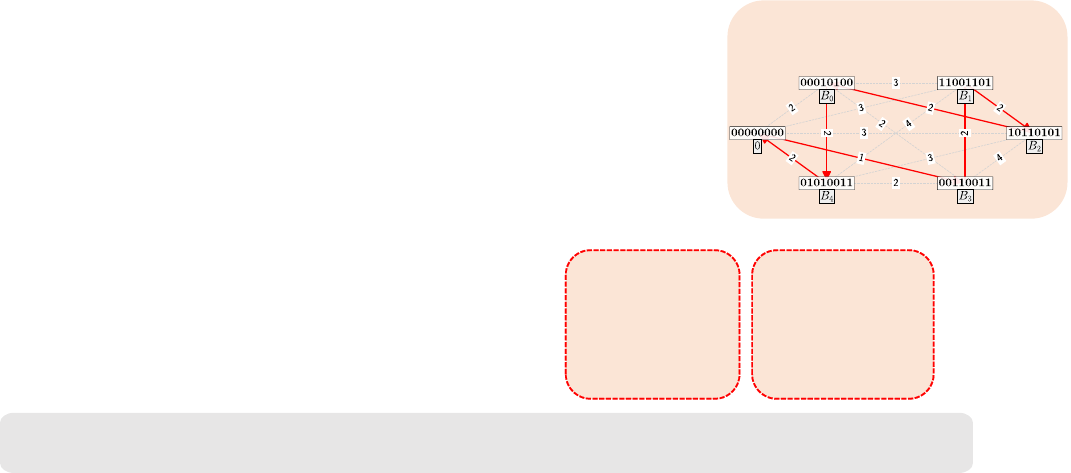}}%
    \put(0.76986856,0.01221805){\color[rgb]{0,0,0}\makebox(0,0)[lt]{\lineheight{1.25}\smash{\begin{tabular}[t]{l}5) encoded state\end{tabular}}}}%
    \put(0,0){\includegraphics[width=\unitlength,page=2]{overview.pdf}}%
    \put(0.45,0.41306715){\color[rgb]{0,0,0}\makebox(0,0)[lt]{\lineheight{1.25}\smash{\begin{tabular}[t]{l}3) construction of\end{tabular}}}}%
    \put(0.71024023,0.41306715){\color[rgb]{0,0,0}\makebox(0,0)[lt]{\lineheight{1.25}\smash{\begin{tabular}[t]{l}4) finding optimal path through\\\hspace{4mm}columns of \end{tabular}}}}%
    \put(0,0){\includegraphics[width=\unitlength,page=3]{overview.pdf}}%
    \put(0.01665004,0.41306715){\color[rgb]{0,0,0}\makebox(0,0)[lt]{\lineheight{1.25}\smash{\begin{tabular}[t]{l}1) data to encode\end{tabular}}}}%
    \put(0,0){\includegraphics[width=\unitlength,page=4]{overview.pdf}}%
    \put(0.21896757,0.41306715){\color[rgb]{0,0,0}\makebox(0,0)[lt]{\lineheight{1.25}\smash{\begin{tabular}[t]{l}2) angle expansion\\\hspace{4mm}to binary matrix\end{tabular}}}}%
    \put(0.45,0.15807993){\color[rgb]{0,0,0}\makebox(0,0)[lt]{\lineheight{1.25}\smash{\begin{tabular}[t]{l}S\end{tabular}}}}%
    \put(0.88273526,0.14074048){\color[rgb]{0,0,0}\makebox(0,0)[lt]{\lineheight{1.25}\smash{\begin{tabular}[t]{l}...\end{tabular}}}}%
    \put(0.45,0.11000137){\color[rgb]{0,0,0}\makebox(0,0)[lt]{\lineheight{1.25}\smash{\begin{tabular}[t]{l}T\end{tabular}}}}%
    \put(0.45,0.08759901){\color[rgb]{0,0,0}\makebox(0,0)[lt]{\lineheight{1.25}\smash{\begin{tabular}[t]{l}F\end{tabular}}}}%
    \put(0,0){\includegraphics[width=\unitlength,page=5]{overview.pdf}}%
    \put(0.49808868,0.15800188){\color[rgb]{0,0,0}\makebox(0,0)[lt]{\lineheight{1.25}\smash{\begin{tabular}[t]{l}\textit{H}\end{tabular}}}}%
    \put(0,0){\includegraphics[width=\unitlength,page=6]{overview.pdf}}%
	\put(0.03, 0.34){$\bm{v} = \begin{pmatrix} \phantom{-}0.422\; \\ \vdots \\ -0.422 \\ \phantom{-}0.141 \\ \phantom{-}0.450 \end{pmatrix}$}
	\put(0.05, 0.186){$\bm{\theta} = \begin{pmatrix} \phantom{-}0.774 \\ \vdots \\ -0.774 \\ \phantom{-}0.202 \\ \phantom{-}1.000 \end{pmatrix}$}
	\put(0.19, 0.253){$\mathbf{B} = \begin{bmatrix}
0 & 1 & 1 & 0 & 0 \\
0 & 1 & 0 & 0 & 1 \\
0 & 0 & 1 & 1 & 0 \\
1 & 0 & 1 & 1 & 1 \\
0 & 1 & 0 & 0 & 0 \\
1 & 1 & 1 & 0 & 0 \\
0 & 0 & 0 & 1 & 1 \\
0 & 1 & 1 & 1 & 1 \end{bmatrix}$}
	\put(0.23, 0.15){$\Delta_{1,2} = (11001101$}
	\put(0.23, 0.128){$\hspace{10mm}\oplus 10110101)$}
	\put(0.48, 0.3885){$\mathcal{W}_{\bm{b}\prime}\ket{\Psi_{\bm b}}=\ket{\Psi_{\bm{b} \oplus \bm{b}\prime}}$}
	\put(0.805, 0.3885){\hspace{4mm}$\mathbf{B}$}
	\put(0.546, 0.149){$\mathcal{W}_{\Delta_{0,1}}$}
	\put(0.715, 0.149){$\mathcal{W}_{\Delta_{1,2}}$}
	\put(0.92, 0.149){$\mathcal{W}_{\Delta_{L,0}}$}
	\put(0.610, 0.089){$R_y(2\pi)$}
	\put(0.782, 0.089){$R_y(\pi/2)$}
	\put(0.034, 0.038){$ \ket{\zeta} \approx \sqrt{\rho} \, ( 0.422\ket{000}_S + \cdots - 0.422\ket{101}_S + 0.141\ket{110}_S + 0.450\ket{111}_S) \, \ket{0}_T\ket{1}_F$}
	\put(0.034, 0.009){$\hspace{8mm}+ \sqrt{1-\rho}|\Psi_{\mathrm{bad}}\rangle \ket{0}_T \ket{0}_F$}
  \end{picture}%
\endgroup%

%% file: controlled_rotations.tex

\[
\begin{quantikz}[wire types={q,q,n,q,q}]
\lstick{$\ket{B_{0,0}}$}
&[-0pt] \ctrl{4} &[-10pt] \qw &[-10pt] \ \ldots\ &[-10pt] \qw &[-10pt] \qw \\
\lstick{$\ket{B_{0,1}}$}
& \qw & \ctrl{3} & \ \ldots\ \qw & \qw& \qw \\
\lstick{\vdots \;\;\;\;}  &  &  & \ \ddots\ & & \\
\lstick{$\ket{B_{0,L-1}}$}& \qw & \qw & \ \ldots\ & \ctrl{1} & \qw \\
\lstick{$ \ket{0}_{F}$} 
& \gate[style={fill=gate_color_crot}]{R_y(\phi_0)} & \gate[style={fill=gate_color_crot}]{R_y(\phi_1)} &  \ \ldots\ & \gate[style={fill=gate_color_crot}]{R_y(\phi_{L-1})} & \qw
\end{quantikz}
\]


%% file: pictures/circuit_latex/algorithm_2_circuit.tex
\begin{quantikz}
\lstick[wires=3]{$\text{S}$} &[-12pt] \gate[wires=3,style={fill=gate_color_H}]{H} &[-10pt]  \gate[wires=4,style={fill=gate_color_W}]{\mathcal{W}_{\vc{\Delta}_{0,1}}} &[-20pt] \qw &[-12pt] \ \ldots\ \qw &[-12pt] \gate[wires=4,style={fill=gate_color_W}]{\mathcal{W}_{\vc{\Delta}_{j,j+1}}} &[-20pt] \qw &[-12pt] \ \ldots\ \qw &[-12pt] \gate[wires=4,style={fill=gate_color_W}]{\mathcal{W}_{\vc{\Delta}_{L,L+1}}} &[-15pt] \qw \\
 & \qw & \qw & \qw & \ \ldots\ \qw & \qw & \qw& \ \ldots\ \qw & \qw & \qw\\
 & \qw & \qw & \qw & \ \ldots\ \qw & \qw & \qw& \ \ldots\ \qw & \qw & \qw\\
\lstick{$\text{T}$} 
    & \qw & \qw & \ctrl{1} & \ \ldots\ \qw & \qw& \ctrl{1} & \ \ldots\ \qw & \qw & \qw\\
\lstick{$\text{F}$} 
    & \qw & \qw & \gate[style={fill=gate_color_crot}]{R_y(\phi_0)}& \ \ldots\ \qw & \qw & \gate[style={fill=gate_color_crot}]{R_y(\phi_j)} & \ \ldots\ \qw & \qw & \qw 
\end{quantikz}